\documentclass[12pt,letterpaper,tightenlines]{revtex4}
\usepackage{amssymb,amsmath}
\usepackage{psfrag}
\usepackage[dvips]{graphicx}
\usepackage{verbatim}
\usepackage{color}
\definecolor{darkblue}{rgb}{0,0,0.5}
\usepackage[dvips, unicode, colorlinks, bookmarks=true, citecolor=darkblue, linkcolor=darkblue, urlcolor=darkblue]{hyperref}

\newcommand{\calA}{\mathcal{A}}

\begin{document}

\begin{abstract}
  A relatively simple method of overcoming the Standard Quantum Limit in the next-generation Advanced LIGO gravitational wave detector is considered. It is based on the quantum variational measurement with a single short (a few tens of meters) filter cavity. Estimates show that this method allows to reduce the radiation pressure noise at low frequencies ($<100\,\mathrm{Hz}$) to the level comparable with or smaller than the low-frequency noises of non-quantum origin (mirrors suspension noise, mirrors thermal noise, and gravity gradients fluctuations). 
\end{abstract}

\title{Quantum variational measurement in the next generation gravitational-wave detectors}

\author{F.Ya.Khalili}

\email{farid@hbar.phys.msu.ru}
\affiliation{Physics Faculty, Moscow State University, Moscow 119992, Russia}


\maketitle

\section{Introduction}

The sensitivity of the first generation large-scale laser interferometric gravitational wave detectors operating now is extremely high. They can detect mechanical displacement as small as $\sim 10^{-16}\,\mathrm{cm}$ \cite{Marx2006}. This sensitivity provides a real chance to detect gravitational waves from astrophysical sources \cite{Abramovici1992, Thorne1995}. However, routine observations of gravitational waves require at least one order of magnitude better sensitivity.

This sensitivity improvement is planned for the second generation detectors, in particular, the Advanced LIGO \cite{WhitePaper1999, Thorne2000, Fritschel2002, AdvLIGOsite}. As a consequence, the Advanced LIGO sensitivity will be close to the Standard Quantum Limit (SQL) \cite{67a1eBr}. This limitation corresponds to the sensitivity level where the meter measurement noise (the shot noise in the optical interferometric position meters case) becomes equal to the meter back action noise (\emph{i.e.} the radiation pressure fluctuations). The first noise is inversely proportional to the optical power and the second one is directly proportional to it.

Several methods of overcoming the SQL have been proposed. One of the most promising is the \emph{quantum variational measurement} \cite{Unruh1982, 87a1eKh, JaekelReynaud1990, Pace1993, 96a2eVyMa}. It uses correlation between the measurement noise and back-action noise, which allows, in principle, to remove the back-action noise component from the meter output. Frequency-independent correlation can be introduced in the optical position meters relatively easy by using homodyne detector with properly adjusted local oscillator phase. However, in this case the back-action suppression is possible in narrow frequency band only.

The method of creating the frequency-dependent noise correlation in large-scale laser interferometric gravitational wave detectors was proposed and analyzed in detail in the article \cite{02a1KiLeMaThVy}. It is based on the use of additional \emph{filter cavities} which introduce frequency-dependent phase shift into the reflected light. These cavities can be placed before the main interferometer (so-called \emph{modified input optics} case) as well as after it (\emph{modified output optics}). In the former case a squeezed quantum state have to be used. In the latter one, it is not necessary but desirable because allows to decrease the required optical power. The comprehensive analysis of this technology application to the signal recycled topology, planned for the Advanced LIGO and used in the operating now smaller GEO-600 gravitational wave detector \cite{Willke2002}, was performed in the articles \cite{Harms2003, Buonanno2004}.

The main technical problem of this method arises due to the requirement that the filter cavities bandwidths should be of the same order of magnitude as the gravitational-wave signal frequency $\Omega\sim10^3\,{\rm s}^{-1}$. Therefore, the filter cavities quality factors have to be as high as $\omega_p/\Omega\sim10^{12}$, where $\omega_p\sim10^{15}\,{\rm s}^{-1}$ is the laser pumping frequency. Therefore, long filter cavities with very high-reflectivity mirrors should be used.  Filter cavities with the same length as the main interferometer cavities (4\,Km), placed in the same vacuum chamber side-by-side with the latter ones, were considered in the article \cite{02a1KiLeMaThVy}. This topology allows, in principle, to obtain sensitivity significantly better than the SQL and probably will be used in the third (post Advanced LIGO) generation of the laser gravitational-wave detectors (see brief discussion on the technical issues of this topology in paper \cite{06a1Kh}). 

In the current article the topology based on the same principle but less ambitious and more simple in implementation is considered. It contains only one relatively short filter cavity with length comparable to the Advanced LIGO auxiliary mode-cleaner cavities: a few tens of meters. We suppose here that the main interferometer parameters values are close to those planned for the Advanced LIGO. In particular, we suppose that:
\begin{itemize}
  \item the optical power circulating in the interferometer arms is equal to the power necessary to reach the SQL, $W = W_\mathrm{SQL}\approx840\,\mathrm{kW}$;
  \item the technical noises: the mirrors thermal noise, the suspensions thermal noise, \emph{etc} are the same as planned for the Advanced LIGO.
  \item the interferometer is tuned in resonance and thus the ``optical springs'' technology \cite{Buonanno2001,Buonanno2002} is not used;
  \item no quantum squeezed states are used
\end{itemize}
(see the brief discussion on the last two items in the Conclusion). It should be noted that the case of short (30 meters long) cavities was considered in paper \cite{Buonanno2004}. However, the authors of this paper followed the original optimization procedure of \cite{02a1KiLeMaThVy} which does not provide very good results for such short cavities. Here we propose another optimization method more suitable for short filter cavities with relatively high optical losses. 

It follows from the planned Advanced LIGO noise budget that the only frequency range where it is possible to increase the sensitivity without the increase of circulating power and/or use of squeezed quantum states, and without reducing the mirrors internal noise, is the low-frequency area $\Omega/2\pi\lesssim100\mathrm{Hz}$, where the sensitivity is limited by the radiation-pressure noise. It is this area that is considered in this article.

In Sec.\,\ref{sec:scheme}, the measurement scheme and the main noise sources are discussed. In Sec.\,\ref{sec:varmeas} a new ``{soft}'' variant of variational measurement optimization is introduced. In Sec.\,\ref{sec:estimates}, the achievable sensitivity is estimated. The main notations and parameters values used in this paper are listed in Table\,\ref{tab:notations}.

\begin{table*}[t]
  \begin{tabular}{|c|c|l|}
    \hline
      Quantity    & Value for estimates                 & Description \\
    \hline 
      $\Omega$    &                                     & Gravitational-wave frequency \\
      $c$         & $3\times10^8\,\mathrm{m/s}$         & Speed of light \\
      $\omega_p$  & $1.77\times10^{15}\,\mathrm{s}^{-1}$& Optical pumping frequency \\
      $M$         & $40\,\mathrm{kg}$                   & Mirror mass \\
      $L$         & $4\,\mathrm{km}$                    & Interferometer arms length \\
      $\gamma$    &                                     & Interferometer half-bandwidth \\
      $W$         & $840\,\mathrm{kW}$                  & Power circulating in each of the arms \\
      $J=\dfrac{8\omega_pW}{McL}$ & $(2\pi\times100\,\mathrm{s}^{-1})^3$ & \\
      $L_f$       &                                     & Filter cavity length \\
      $T_f^2$     &                                     & Filter cavity input mirror transmittance \\
      $A_f^2$     &                                     & Filter cavity losses per bounce \\
      $\gamma_{f\,\mathrm{load}}=\dfrac{cT_f^2}{4L}$ &  & \\[1ex]
      $\gamma_{f\,\mathrm{loss}}=\dfrac{cA_f^2}{4L}$ &  & \\
      $\gamma_f=\gamma_{f\,\mathrm{load}}+\gamma_{f\,\mathrm{loss}}$ & & Filter cavity half-bandwidth\\
      $\delta_f$  &                                     & Filter cavity detuning \\
    \hline
  \end{tabular}
  \caption{Main notations used in this paper.}\label{tab:notations}
\end{table*}

\section{The scheme}\label{sec:scheme}

\begin{figure*}
  \psfrag{m}[cb][lb]{$M$}\psfrag{m1}[lc][lb]{$M$}
  \psfrag{W}[cb][lb]{$W$}\psfrag{W1}[lc][lb]{$W$}\psfrag{WE}[cb][lb]{$2W$}
  \psfrag{C}[rb][lb]{\sf Circulator}
  \psfrag{D}[rc][lb]{\parbox{0.15\textwidth}{\sf To homodyne\\detector}}
  \psfrag{F}[lc][lb]{\sf Filter cavity}
  \psfrag{FM}[cb][lb]{\sf Fixed}
  \includegraphics[width=0.48\textwidth]{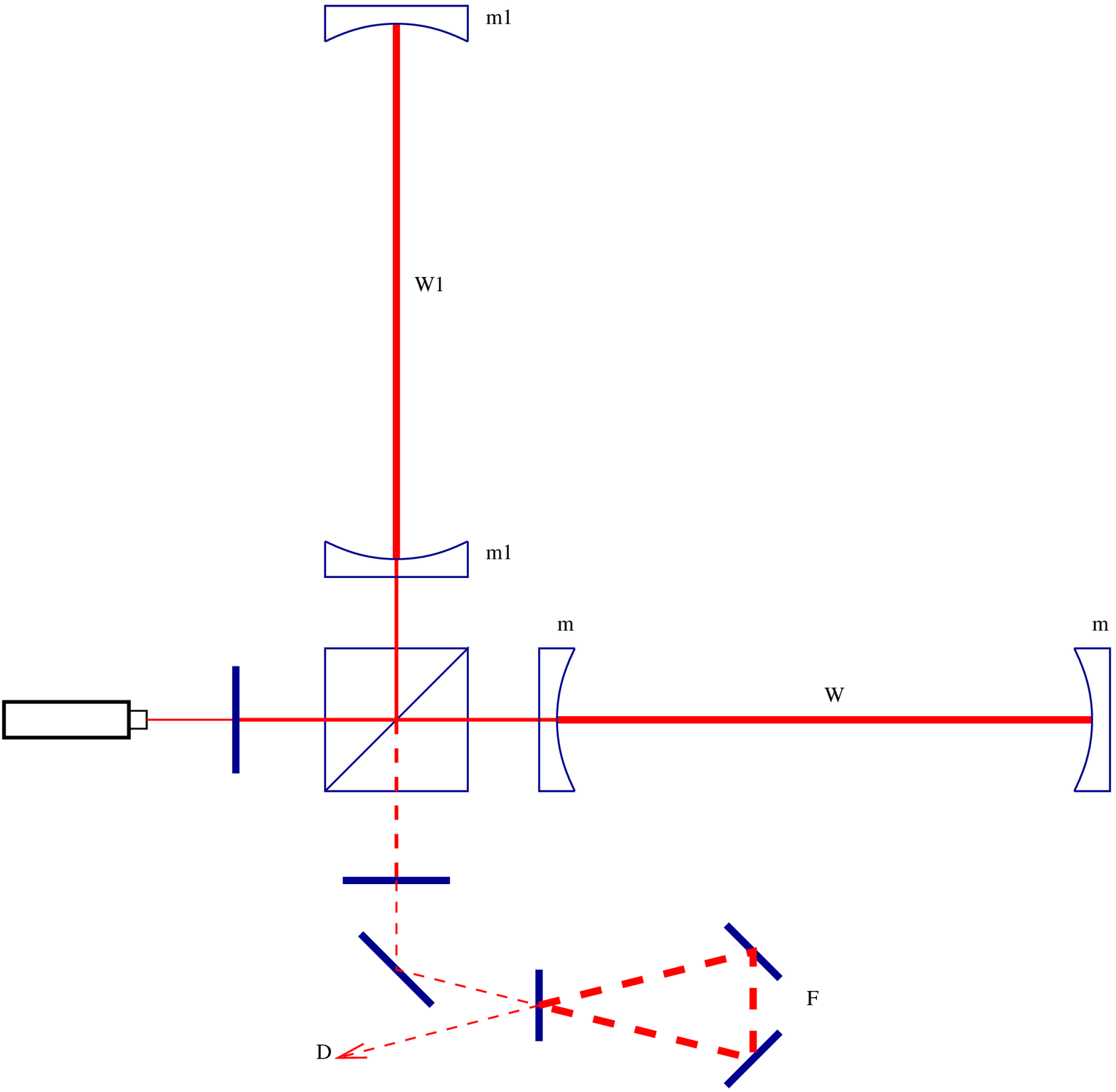}\hfill
  \includegraphics[width=0.48\textwidth]{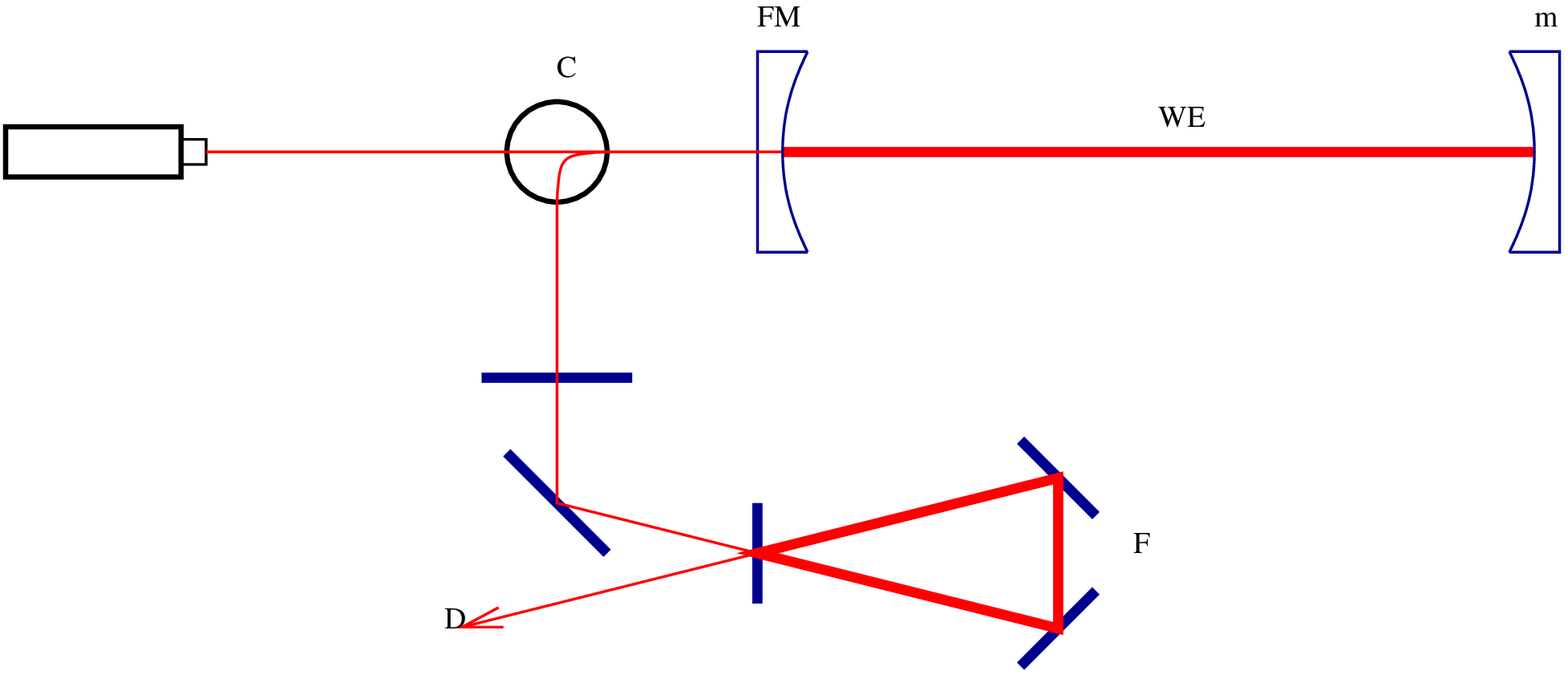}
  \caption{Left: the signal/power recycled Advanced LIGO interferometer with additional filter cavity; right: simplified equivalent scheme.}\label{fig:scheme}
\end{figure*}

The scheme considered in this article is shown in Fig.\ref{fig:scheme}(left). Basically, it is signal/power recycled Advanced LIGO interferometer topology with additional filter cavity in the output port. Variant with ring filter cavity is shown. An ordinary Fabri-Perot cavity can be also used; however, some circulator is necessary in this case to separate the filter cavity output beam from the input one. On the other hand, only one end mirror is required instead of two in the latter case, which allows to decrease the cavity internal losses.

It was shown in article \cite{Buonanno2003}, that the Advanced LIGO topology can be mapped to a simple Fabri-Perot cavity based optical position meter, with fixed input mirror and movable end one. The end mirror mass should be equal to the mass of each of four Advanced LIGO mirrors, the cavity length should be equal to the Advanced LIGO arms length, and the circulating optical power should be twice as high as the circulating power in each of the Advanced  LIGO arms [see Fig.\ref{fig:scheme}(right)]. It is this equivalent scheme that will be used for calculations below.

The scheme quantum noise includes components introduced by the optical losses in the main interferometer, the optical losses in the filter cavity, and the homodyne detector non-unity quantum efficiency. The first and the third noise sources can be characterized together by the unified quantum efficiency
\begin{equation}
  \eta = \frac{\eta_\mathrm{det}}{1+\dfrac{cA^2}{4L\gamma}} \,,
\end{equation}
where $A^2$ is the Interferometer optical losses per bounce. In the estimates below, two values of $\eta$ will be used: $\eta=0.9$ (equal to one planned for the Advanced LIGO) and $\eta=0.99$. It have to be noted, that in the Advanced LIGO case, $cA^2/4L\gamma\ll10^{-3}$, and the value of $\eta$ is defined primarily by $\eta_\mathrm{det}$.

The filter cavity losses appear in all equations through the specific loss factor $A_f^2/L_f$. In the estimates, the following values of this parameter will be used: $5\times10^{-7}\,\mathrm{m}^{-1}$ (for example, 20\,m filter cavity with $A_f^2=10^{-5}$) and $10^{-7}\,\mathrm{m}^{-1}$.

\section{Soft variational measurement}\label{sec:varmeas}

The sum quantum noise of the scheme in Fig.\ref{fig:scheme} has been calculated in paper \cite{06a1Kh}. 
In the current section, we use simplified expressions where only linear in the filter cavity loss factor terms are kept. This approximation holds rather well down to frequencies $\Omega\sim(2\div3)\gamma_{f\,\mathrm{loss}}$ and make the analysis much more transparent.

If the interferometer is tuned in resonance, then the sum quantum noise can be presented as follows:
\begin{equation}\label{xi2_1}
  \xi^2(\Omega) \equiv \frac{S^h(\Omega)}{S^h_\mathrm{SQL}(\Omega)} 
  = \frac{\Omega^2(\Omega^2+\gamma^2)}{4J\gamma\eta_\Sigma(\Omega)\cos^2\phi_\Sigma(\Omega)}
    - \tan\phi_\Sigma(\Omega) + \frac{J\gamma}{\Omega^2(\Omega^2+\gamma^2)} \,,
\end{equation} 
where $S^h(\Omega)$ is spectral density of the equivalent strain noise, 
\begin{equation} S^h_\mathrm{SQL}(\Omega) = \frac{8\hbar}{ML^2\Omega^2} \end{equation} 
is the spectral density corresponding to the SQL,
\begin{equation} \phi_\Sigma(\Omega) =  \phi + \phi_f(\Omega) \,, \end{equation} 
$\phi$ is the local oscillator phase,
\begin{equation} 
  \phi_f(\Omega) 
    = \arctan\frac{2\gamma_{f\,\mathrm{load}}\delta_f}{\Omega^2+\gamma_{f\,\mathrm{load}}^2-\delta_f^2} 
\end{equation} 
is the phase shift in filter cavity, and 
\begin{equation} \eta_\Sigma(\Omega) = \frac{\eta}{1 + \calA_f(\Omega)} \end{equation} 
is the quantum efficiency of the scheme, where
\begin{equation} 
  \calA_f(\Omega) = \frac{
      4\gamma_{f\,\mathrm{loss}}\gamma_{f\,\mathrm{load}}(\Omega^2+\gamma_{f\,\mathrm{load}}^2+\delta_f^2)
    }{
      \Omega^4 + 2\Omega^2(\gamma_{f\,\mathrm{load}}^2-\delta_f^2)
      + (\gamma_{f\,\mathrm{load}}^2+\delta_f^2)^2
    } 
\end{equation}
is the effective loss factors of the filter cavity.

It is convenient to separate in Eq.\,(\ref{xi2_1}) terms of different origin:
\begin{equation}\label{xi2_2}
  \xi^2(\Omega) = \xi_\mathrm{SN}^2(\Omega) + \xi_\mathrm{res}^2(\Omega) + \xi_\mathrm{loss}^2(\Omega)\,,
\end{equation} 
where
\begin{subequations}\label{xi2_comps}
\begin{equation}\label{xi2_SN}
  \xi_\mathrm{SN}^2(\Omega) = \frac{\Omega^2(\Omega^2+\gamma^2)}{4J\gamma\eta_\Sigma(\Omega)}
\end{equation} 
is the component created by the shot noise,
\begin{equation}\label{xi2_res}
  \xi_\mathrm{res}^2(\Omega) = \xi_\mathrm{SN}^2(\Omega)
    \left[\tan\phi_\Sigma(\Omega) - \frac{1}{2\xi_\mathrm{SN}^2(\Omega)}\right]^2
\end{equation} 
is the residual part of the back-action noise, and
\begin{equation}\label{xi2_loss}
  \xi_\mathrm{loss}^2(\Omega) = \frac{J\gamma[1-\eta_\Sigma(\Omega)]}{\Omega^2(\Omega^2+\gamma^2)}
\end{equation} 
\end{subequations}
is the component created by optical losses.

The standard variational measurement approach is to eliminate completely the back-action noise by setting 
\begin{equation}\label{exact_VM}
  \tan\phi_\Sigma(\Omega) -\frac{1}{2\xi_\mathrm{SN}^2(\Omega)} \equiv 0\,.
\end{equation} 
With only one filter cavity, this equality can not be fulfilled exactly for all frequencies $\Omega$. Moreover, in the presence of optical losses, two filter cavities are also insufficient for it. On the other hand, in case of short filter cavity, the exact fulfillment of condition (\ref{exact_VM}) does not allow to obtain arbitrary high sensitivity, anyway, due to the losses term  (\ref{xi2_loss}). 

At the same time, the term  (\ref{xi2_res}) can be reduced significantly and made smaller than the other two terms in Eq.\,(\ref{xi2_2}) even when only one filter cavity is used. Condition (\ref{exact_VM}) should be fulfilled approximately in this case:
\begin{equation}\label{soft_VM}
  \tan\phi_\Sigma(\Omega) - \frac{1}{2\xi_\mathrm{SN}^2(\Omega)} \approx 0 \,.
\end{equation} 

It is evident that this \emph{soft variational measurement} optimization can be performed in many different ways, depending on the desirable shape of the resulting noise spectral density. However, in the  particular case considered here, taking into account the constraints listed in the Introduction, the parameters choice is rather unique.

Let us start with the high-frequency area, $\Omega\gg\gamma$, where 
\begin{align}
  \xi_\mathrm{loss}(\Omega) &\to 0\,, 
  & \xi_\mathrm{res}^2(\Omega) \approx \xi_\mathrm{SN}^2(\Omega)\tan^2\phi \,,
\end{align}
and
\begin{equation}
  \xi^2(\Omega) \approx \frac{\xi_\mathrm{SN}^2(\Omega)}{\cos^2\phi} \,.
\end{equation} 
Therefore, in order to keep this noise as small as possible, there should be
\begin{equation}\label{phi0} \phi=0 \,. \end{equation} 
It should be noted, that in ordinary meters (\emph{i.e.}, without variational measurement), using $\phi\ne0$, it is possible to obtain some sensitivity gain at low ($\Omega<\gamma$) or medium ($\Omega\sim\gamma$) frequencies at the cost of increased high-frequency noise. With  variational measurement this trade-off is possible too, but in this case the gain is small compared to the gain provided by variational measurement itself. Therefore, the only case of $\phi=0$ is considered here.

At low frequencies $\xi_\mathrm{res}^2$ is approximately equal to
\begin{equation}
  \xi_\mathrm{res}^2(\Omega)\Bigr|_\mathrm{\Omega\to0}
  \approx\frac{\Omega^2\gamma}{J\eta_\Sigma(0)}\left[
    \frac{\gamma_{f\,\mathrm{load}}\delta_f}{\Omega^2+\gamma_{f\,\mathrm{load}}^2-\delta_f^2} 
    - \frac{J\eta_{\Sigma}(0)}{\Omega^2\gamma}
  \right]^2 \,.
\end{equation} 
This expression is equal to zero, if
\begin{align}
  \gamma_{f\,\mathrm{load}} &= \delta_f \,, & 
  \gamma_{f\,\mathrm{load}}\delta_f = \frac{J\eta_\Sigma(0)}{\gamma} \,.
\end{align}
The last equations specify the filter cavity parameters:
\begin{equation}\label{gfdf_opt}
  \gamma_{f\,\mathrm{load}} = \delta_f 
   = \sqrt{\frac{J\eta}{\gamma} + \gamma_{f\,\mathrm{loss}}^2} - \gamma_{f\,\mathrm{loss}} 
   \approx \sqrt{\frac{J\eta}{\gamma}} - \gamma_{f\,\mathrm{loss}}\,.
\end{equation} 

The only free parameter remaining is the interferometer half-bandwidth $\gamma$. Typically, it is supposed for wide band configurations like the one considered here, that $\gamma\approx J^{1/3}$. However, asymptotic values of $\xi(\Omega)$ at both low and high frequencies decrease with $\gamma$ increase. The larger values of $\gamma\sim1000\div3000\,\mathrm{s}^{-1}$ also provide significantly larger values of the signal-to-noise ratio for neutron star - neutron star inspiral waveforms (see next section).

\begin{figure}
  \psfrag{f}[ct][ct]{$\Omega/2\pi$}
  \psfrag{xi2}[cc][cc]{$\xi^2(\Omega)=\dfrac{S^h(\Omega)}{S^h_\mathrm{SQL}(\Omega)}$}
  \psfrag{loss}[lb][lb]{$\xi_\mathrm{loss}^2(\Omega)$}
  \psfrag{res}[rb][lb]{$\xi_\mathrm{res}^2(\Omega)$}
  \psfrag{SN}[lb][lb]{$\xi_\mathrm{SN}^2(\Omega)$}
  \includegraphics[width=0.5\textwidth,height=0.5\textwidth]{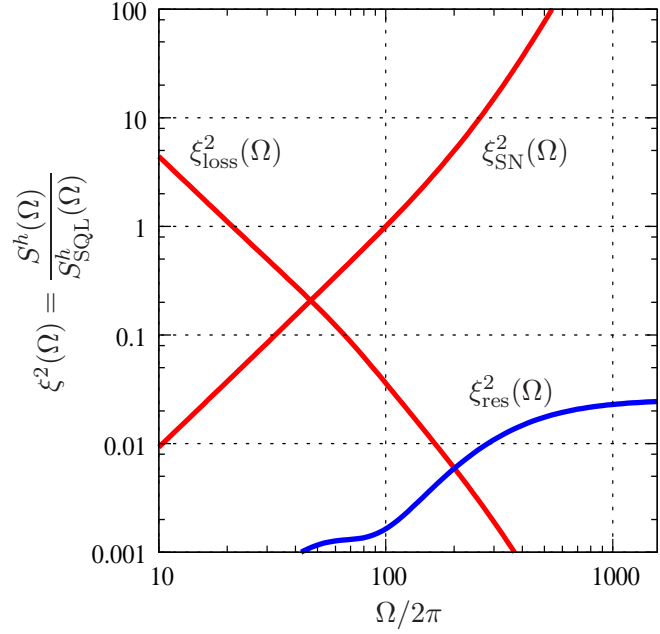}
  \caption{Plots of the noise components $\xi_\mathrm{loss}^2(\Omega)$, $\xi_\mathrm{res}^2(\Omega)$, and $\xi_\mathrm{SN}^2(\Omega)$. Filter cavity losses are equal to $A_f^2/L_f=5\times10^{-7}\,\mathrm{m}^{-1}$ and $\eta=0.9$.}\label{fig:components} 
\end{figure} 

In order to demonstrate the back action noise suppression level in the optimization procedure considered above, the noise components $\xi_\mathrm{loss}^2(\Omega), \xi_\mathrm{res}^2(\Omega)$, and $\xi_\mathrm{SN}^2(\Omega)$ are plotted separately in Fig.\,\ref{fig:components}, for $A_f^2/L_f=1\times10^{-7}\,\mathrm{m}^{-1}$ and $\gamma = 2000\,\mathrm{s}^{-1}$. It follows from this plot, that the residual back action noise is indeed much smaller than at least one of two other noise components.

\section{The sensitivity}\label{sec:estimates}

The exact expression for the quantum noise is the following:
\begin{multline}
  \xi^2(\Omega)
    = \frac{\Omega^2(\Omega^2+\gamma^2)}{4J\gamma\eta}\,
        \frac{\Omega^4 + 2\Omega^2(\gamma_f^2-\delta_f^2) + (\gamma_f+\delta_f^2)^2}
          {(\Omega^2 + \gamma_{f\,\mathrm{load}}^2-\delta_f^2)^2 + 4\Omega^2\gamma_{f\,\mathrm{loss}}^2}
         \\ 
      - \frac{2\gamma_{f\,\mathrm{load}}\delta_f(\Omega^2 + \gamma_{f\,\mathrm{load}}^2-\delta_f^2)}
          {(\Omega^2 + \gamma_{f\,\mathrm{load}}^2-\delta_f^2)^2 + 4\Omega^2\gamma_{f\,\mathrm{loss}}^2}
      + \frac{J\gamma}{\Omega^2(\Omega^2+\gamma^2)} \,.
\end{multline}
Due to its sophisticated dependence on the filter cavity parameters and the observation frequency $\Omega$, universal optimization similar to one considered in the previous section it is impossible in this general case. Instead, some frequency independent sensitivity measure have to be chosen and optimized. The noises of non-quantum origin also have to be taken into account in this optimization.

Following article \cite{Buonanno2004}, the signal-to-noise ratio for neutron star - neutron star (NSNS) inspiral waveforms:
\begin{equation}\label{SNR_NSNS}
  \mathrm{SNR}_\mathrm{NSNS} 
    = A\int_{f_C}^{f_\mathrm{ISCO}}\frac{f^{-7/3}}{S^h(2\pi f) + S^h_\text{no-q}(2\pi f)}\,df
\end{equation} 
will be used as such a measure in this paper. Here $f_c=10\,\mathrm{Hz}$ is the gravitational-wave detector low-frequency cut-off, $f_\mathrm{ISCO}=1570\,\mathrm{Hz}$ is the gravitational-wave frequency corresponding to the Innermost Stable Circular Orbit of a Schwarzchild black hole with mass equal to $2\times1.4$ solar masses, and $A$ is a numeric factor which does not depend on the scheme specific parameters. $S^h_\text{no-q}(\Omega)$ is the sum spectral density of the most important non-quantum noises: the gravitational-wave detector mirrors thermal noise, the mirrors suspension thermal noise, and the Newtonian gravity gradient noise. In the estimates here, the value of $S^h_\text{no-q}(\Omega)$ calculated by means of the {\sf bench} program \cite{benchsite} will be used. The interferometer parameters coded into this program correspond to the parameters set and the technology level planned for the Advanced LIGO.

For convenience we normalize this signal-to-noise ratio by the one corresponding to the conventional (SQL-limited) gravitational-wave detector with the half-bandwidth $\gamma$ equal to $2\pi\times100\,\mathrm{s}^{-1}$:
\begin{equation}\label{SNRR_NSNS}
  \mathrm{SNRR}_\mathrm{NSNS} = \frac{
      \displaystyle\int_{f_C}^{f_\mathrm{ISCO}}
        \frac{f^{-7/3}}{S^h(2\pi f) + S^h_\text{no-q}(2\pi f)}\,df
    }{
      \displaystyle\int_{f_C}^{f_\mathrm{ISCO}}
        \frac{f^{-7/3}}{S^h_\mathrm{conv}(2\pi f)|_{\gamma=2\pi\times100\,\mathrm{s}^{-1}} 
        + S^h_\text{no-q}(2\pi f)}\,df
    } \,,
\end{equation} 
where
\begin{equation}
  S^h_\mathrm{conv}(\Omega) =  \frac{8\hbar}{ML^2\Omega^2}\left[
    \frac{\Omega^2(\Omega^2+\gamma^2)}{4J\gamma} + \frac{J\gamma}{\Omega^2(\Omega^2+\gamma^2)}
  \right] \,.
\end{equation} 

\begin{figure}
  \psfrag{g}[ct][ct]{$\gamma$}
  \psfrag{SNR}[cc][cc]{$\mathrm{SNRR}_\mathrm{NSNS}$}
  \includegraphics[width=0.49\textwidth]{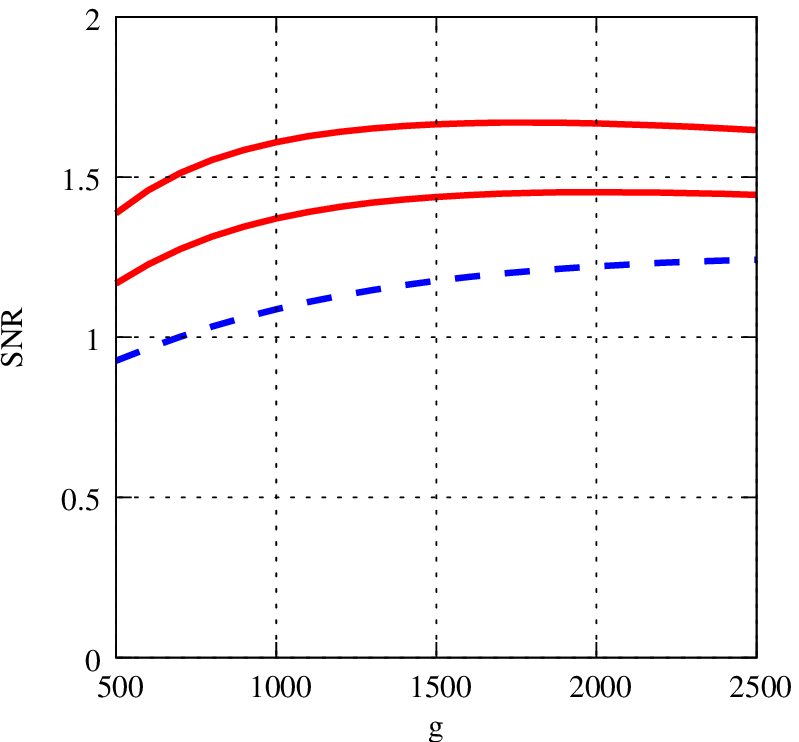}\hfill
  \includegraphics[width=0.49\textwidth]{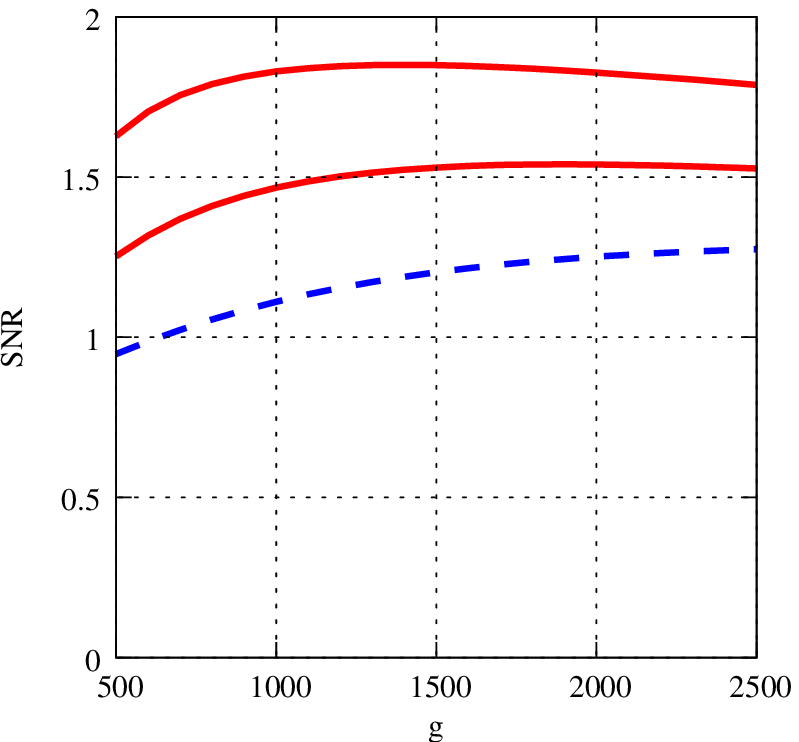}
  \caption{Plots of $\mathrm{SNRR}_\mathrm{NSNS}$ as a function of $\gamma$ for the single filter cavity variational interferometer (solid lines) and for the conventional (SQL-limited) one (dashed lines). Left pane: $\eta=0.9$; right pane: $\eta=0.99$. On each pane, upper solid line corresponds to $A_f^2/L_f=1\times10^{-7}\,\mathrm{m}^{-1}$ and lower one --- to $A_f^2/L_f=5\times10^{-7}\,\mathrm{m}^{-1}$.}\label{fig:SNRR_gamma}
\end{figure}

Expression (\ref{SNRR_NSNS}) contains three free parameters: the filter cavity half-bandwidth $\gamma_{f\,\mathrm{load}}$ and detuning $\delta_f$, and the main interferometer half-bandwidth $\gamma$. We optimize numerically this function with respect to the first two parameters, obtaining thus a function of one argument $\gamma$. This function is plotted in Fig.\,\ref{fig:SNRR_gamma} for the values of losses in the main interferometer and in the filter cavity noted in Sec.\,\ref{sec:scheme}. For comparison function (\ref{SNRR_NSNS}) for the conventional interferometer (\emph{i.e.}, the one with $S^h=S^h_\mathrm{conv}$) is also presented.

In Table \ref{tab:results}, the maximal values of $\mathrm{SNRR}_\mathrm{NSNS}$ are listed for the same combinations of the losses parameters, together with the optimal values of $\gamma_{f\,\mathrm{load}}$, $\delta_f$, and $\gamma$. The maximal values of $\mathrm{SNRR}_\mathrm{NSNS}$ for the conventional interferometer as well as the ones for the signal recycled Advanced LIGO topology are also included into the table. In the latter case, the ``{default}'' parameters set coded into the {\sf bench} program is used: the arm cavities half-bandwidth $93.75\,\mathrm{s}^{-1}$, signal recycling mirror transmittance $\rho=0.06$, signal recycling cavity detuning $(\pi-0.09)/2$, and the homodyne phase $\zeta=\pi/2$ (the latter one differ from the homodyne phase $\phi$ used elsewhere in this paper by the phase shift in the signal recycling cavity). Using the ``{scaling low}'' theorem of paper \cite{Buonanno2003}, these low-level parameters can be translated to the shown in Table \ref{tab:results} half-bandwidth $\gamma$ and detuning $\delta$ of the equivalent Fabri-Perot cavity. 

\begin{table}
  \begin{tabular}{|l|c|l|}
    \hline
    Topology \& losses & $\mathrm{SNRR}_\mathrm{NSNS}$ & Optimal parameters \\
    \hline
    Conventional, $\eta=0.9$ & 1.25 & $\gamma\approx3000\,\mathrm{s^{-1}}$ \\
    Conventional, $\eta=0.99$ & 1.28 & $\gamma\approx3000\,\mathrm{s^{-1}}$ \\
    Advanced LIGO signal recycled, $\eta=0.9$ & 1.43 & 
      $\gamma=641\,\mathrm{s}^{-1}$, $\delta=1860\,\mathrm{s}^{-1}$ \\
    Advanced LIGO signal recycled, $\eta=0.99$ & 1.46 & 
      $\gamma=641\,\mathrm{s}^{-1}$, $\delta=1860\,\mathrm{s}^{-1}$ \\
    Variational, $\eta=0.9$, $A_f^2/L_f=5\times10^{-7}\,\mathrm{m}^{-1}$ & 
      1.45 &
       $\gamma\approx2000\,\mathrm{s^{-1}}$,
       $\gamma_{f\,\mathrm{load}}\approx367\,\mathrm{s^{-1}}$, $\delta_f\approx337\,\mathrm{s^{-1}}$ \\
    Variational, $\eta=0.9$, $A_f^2/L_f=1\times10^{-7}\,\mathrm{m}^{-1}$ & 
      1.67 &
      $\gamma\approx1800\,\mathrm{s^{-1}}$,
      $\gamma_{f\,\mathrm{load}}\approx347\,\mathrm{s^{-1}}$, $\delta_f\approx346\,\mathrm{s^{-1}}$ \\
    Variational, $\eta=0.99$, $A_f^2/L_f=5\times10^{-7}\,\mathrm{m}^{-1}$ & 
      1.54 &
      $\gamma\approx1900\,\mathrm{s^{-1}}$,
      $\gamma_{f\,\mathrm{load}}\approx398\,\mathrm{s^{-1}}$, $\delta_f\approx370\,\mathrm{s^{-1}}$ \\
    Variational, $\eta=0.99$, $A_f^2/L_f=1\times10^{-7}\,\mathrm{m}^{-1}$ & 
      1.85 &
      $\gamma\approx1300\,\mathrm{s^{-1}}$,
      $\gamma_{f\,\mathrm{load}}\approx409\,\mathrm{s^{-1}}$, $\delta_f\approx409\,\mathrm{s^{-1}}$ \\
    \hline 
  \end{tabular}
  \caption{The values of $\mathrm{SNRR}_\mathrm{NSNS}$ for conventional (SQL-limited), signal recycled and variational interferometers.}\label{tab:results}
\end{table}

\begin{figure}
  \psfrag{f}[ct][ct]{$\Omega/2\pi$}
  \psfrag{S}[cc][cc]{$\sqrt{S^h(\Omega)}$}
  \includegraphics[height=0.5\textwidth]{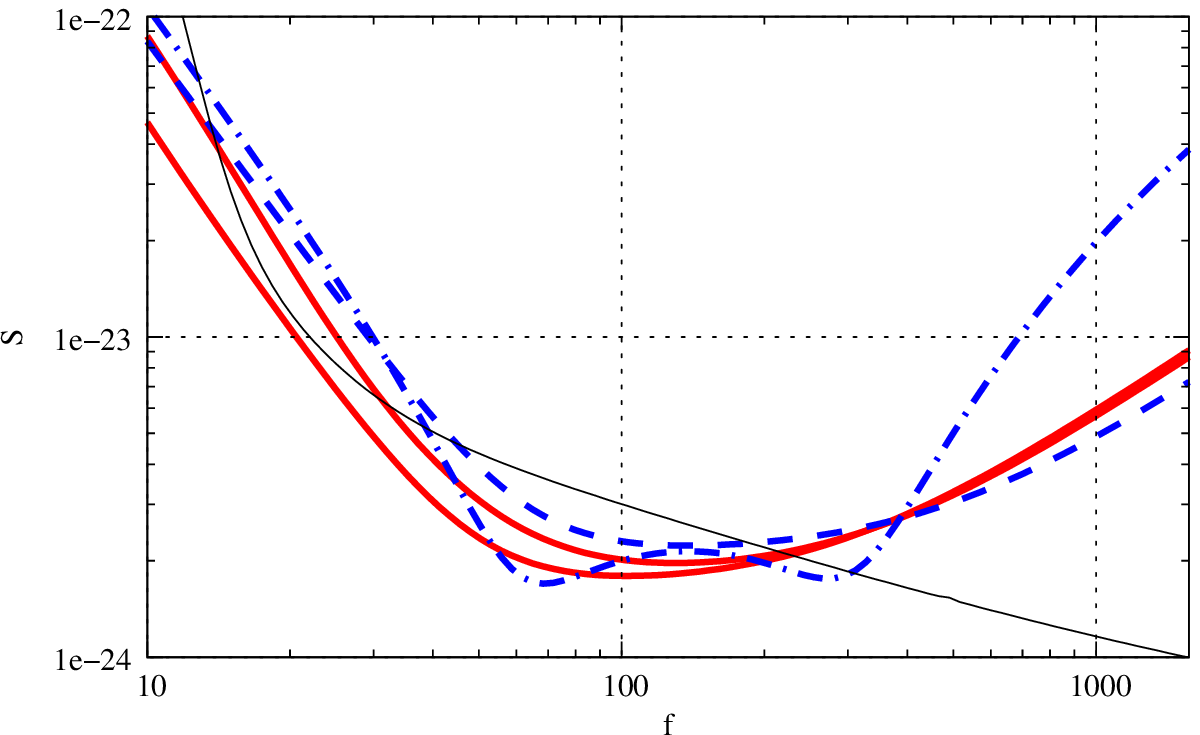}
  \includegraphics[height=0.5\textwidth]{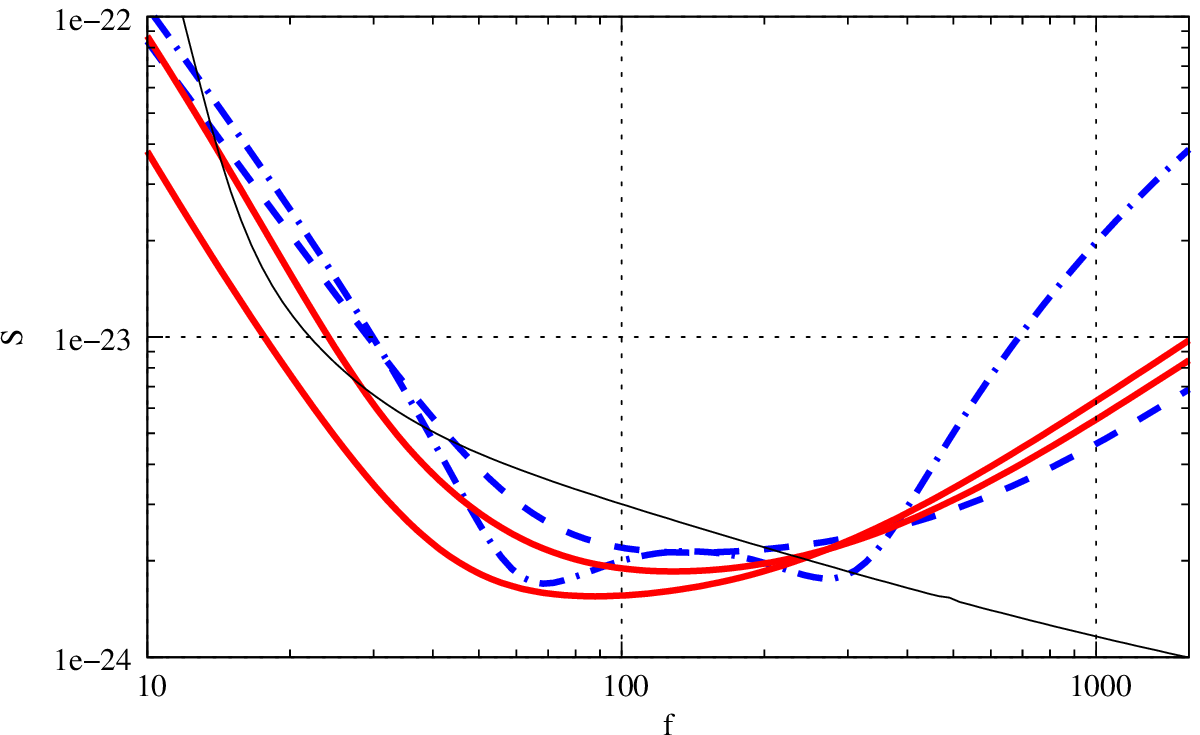}
  \caption{Quantum noise curves for the schemes listed in Table \ref{tab:results}: the single filter cavity variational interferometer (thick solid lines; upper line --- $A_f^2/L_f=5\times10^{-7}\,\mathrm{m}^{-1}$, lower one --- $A_f^2/L_f=1\times10^{-7}\,\mathrm{m}^{-1}$); conventional SQL-limited interferometer (dashed line); signal recycled (Advanced LIGO) interferometer (dash-dotted line). Thin solid line --- the sum non-quantum noise (mirrors thermal + mirrors suspension thermal + gravity gradients). Upper pane: $\eta=0.9$, lower one: $\eta=0.99$.}\label{fig:S_sep}
\end{figure}

The sensitivity frequency dependence of the schemes listed in Table\,\ref{tab:results} is shown in Fig.\,\ref{fig:S_sep}, where square roots of the corresponding quantum noise spectral densities $\sqrt{S^h(\Omega)}$ are plotted using the parameters sets of Table\,\ref{tab:results}, together with the square root of the sum non-quantum noise $S^h_\text{no-q}(2\pi f)$.

Figs.\,\ref{fig:SNRR_gamma}, \ref{fig:S_sep} and Table \ref{tab:results} allow to compare directly quantum noise of the scheme considered here with the main noises of non-quantum origin predicted for the Advanced LIGO, and also with the equivalent strain noise of the ordinary wide-band SQL-limited meter. The direct comparison with the quantum noise of the signal recycled Advanced LIGO topology is, strictly speaking, incorrect, because this topology involves another technique of overcoming the SQL, based on the optical rigidity, with the goal to increase the SNR for neutron star-neutron star inspiral events. As a result, the quantum noise is suppressed at medium frequencies ($\sim100\,\mathrm{Hz}$), while rises at higher frequencies. It could be noted, however, that the variational scheme considered here has the same signal-to-noise ratio for this type of signals even for the worst of the considered losses combinations ($\eta=0.9$, $A_f^2/L_f=1\times10^{-7}\,\mathrm{m}^{-1}$), while providing a more flat  broadband noise curve. 

\section{Conclusion}

It follows from the estimates made in this paper that using variational measurement with single relatively short filter cavity, it is possible to reduce the back action noise in the Advanced LIGO interferometer to the level comparable to or smaller than the low-frequency noises of non-quantum origin: mirrors suspension noise, mirrors internal thermal noise, and gravity gradients fluctuations. The minimal reasonable filter cavity length for the best mirrors available (with losses per bounce $\sim 10^{-5}$) is about $20\,\mathrm{ m}$. In this case, about 2-fold  increase of the Advanced LIGO sensitivity at low frequencies is feasible. Better mirrors, with losses per bounce $\lesssim 5\times10^{-6}$ and/or longer filter cavity would be able to virtually remove the back-action noise from the Advanced LIGO noise budget.

It is evident, that the scheme considered here can be combined with other methods, which also allows to increase the interferometric gravitational wave detectors sensitivity without significant modifications of the topology and without the increase of optical power. In particular, using the the optical rigidity \cite{99a1BrKh, Buonanno2001, Buonanno2002}, it is possible to reshape the noise spectral dependence, extending the low frequency sensitivity gain to the medium frequencies range. 

Another promising option is the squeezed vacuum injection into the interferometer dark port \cite{Caves1981}. It allows to decrease the shot noise at the cost of increased radiation pressure noise (for the same value of the mean optical power). The radiation pressure noise, in turn, can be reduced at low frequencies by using variational measurement of the type considered here. In paper \cite{Corbitt2004-3} the method of generation of squeezed states with frequency-dependent \emph{amplitude} of squeezing, which could provide an additional suppression of the radiation-pressure noise, was proposed. Taking into account the recent achievements in preparation of squeezed quantum states at low frequencies ($10\div1000\,\mathrm{Hz}$) \cite{McKenzie2004, Valbruch2006}, it possible to hope that this combination could provide the sensitivity gain of $\sim2\div3$ within the entire Advanced LIGO frequency band. 

\acknowledgments

This work was supported by the NSF and Caltech grant PHY-0353775 and by Russian government grant NSh-5178.2006.2.

The author is grateful to V.Braginsky, Y.Chen, S.Danilishin, N.Mavalvala, K.Strain, and S.Vyatchanin for stimulating discussions and useful remarks. The author is grateful also to Y.Chen for the remark about the ``{slack}'' low-frequency area in the Advanced LIGO noise budget which stimulated this paper writing.


\end{document}